\documentclass{elsart}
\usepackage{natbib}
\usepackage{amssymb}

\begin{document}

\begin{frontmatter}



\title{Creation of Quark-gluon Plasma in Celestial Laboratories}

\author{R. K. Thakur}

\address{*Retired Professor of Physics, School of Studies in
 Physics \\ Pt.Ravishakar Shukla University, Raipur, India \\
21 College Road, Choube Colony, Raipur-492001, India
}
\ead{rkthakur0516@yahoo.com}
\begin{abstract}
 It is shown that a gravitationally collapsing
 black hole acts as an ultrahigh energy particle
 accelerator that can accelerate particles to energies 
inconceivable in any terrestrial particle accelerator, and 
that when the energy E of the particles comprising the 
matter in the black hole is $ \sim 10^{2} $ GeV  or more,or 
equivalently the temperature T is $ \sim 10^{15}$ K  or 
more, the entire matter in the black hole will be in the 
form of quark-gluon plasma permeated by leptons.

\end{abstract}


\end{frontmatter}

\section{Introduction}
Efforts are being made to create quark-gluon
plasma (QGP)in terrestrial laboratories. A report released
by CERN, the European Organization for Nuclear Research, 
at Geneva, on February 10, 2000 said, ``A series of 
experiments using CERN's lead beam have presented 
compelling evidence for the existence of a new state of 
matter 20 times denser than nuclear matter, in which 
quarks instead of being bound up into more complex 
particles such as protons and neutrons, are liberated to 
roam freely``. By smashing together lead ions at CERN's 
accelerator at temperatures 100,000 times as hot as sun's 
centre, i.e. at temperatures 
$ T \sim 1.5 \times 10^{12}$ K, and energy densities never 
before reached in laboratory experiments, a team of 350 
scientists from institutes in 20 countries succeeded in 
isolating quarks from more complex particles, e.g. protons 
and neutrons. However, the evidence of creation QGP at CERN
is indirect,involving detection of particles produced when 
QGP changes back to hadrons. The production of these 
particles can be  explained alternatively without having to
have QGP. Therefore, the evidence of the creation of QGP at 
CERN is not enough and conclusive. In view of this CERN 
will start a new experiment, ALICE (A Large Ion Collider 
Experiment), at much higher energies available at its LHC 
(Large Hadron Collider). First collisions in the LHC will 
occur in November 2007. A two months run in 2007, with beams
colliding at an energy of 0.9 TeV, will give the accelerator
and detector teams the opportunity to run-in their 
equipment, ready for a run at the full collision energy of 
14 Tev to start in spring 2008.

In the meantime, the focus of research on QGP has 
shifted to the Relativistic Heavy Ion Collider (RHIC), the 
world's newest and largest particle accelerator for nuclear
research, at Brookhaven National Laboratory (BNL) in Upton,
New York. RHIC's goal is to create and study QGP by head-on
collisions of two beams of gold ions at energies 10 times 
those of CERN's programme, which ought to produce QGP with 
higher temperature and longer life time thereby allowing 
much clear and direct observation. The programme at RHIC 
started in June 2000.Researchers at RHIC generated 
thousands of head-on collisions between gold ions at 
energies of 130 GeV creating fireballs of matter having 
density hundred times greater than that of the nuclear 
matter and temperature $\sim 2 \times 10^{12} $ K (175 MeV
in the energy scale). Fireballs were of size 
$ \sim 5 $ femtometre which lasted a few times $ 10^{-24} $
second. All the four detector systems, viz., STAR, PHENIX, 
BRAHMS, PHOBOS, detected ``jet quenching`` and suppression
of ``leading particles``, highly energetic individual 
particles that emerge from the nuclear fireballs in 
gold-gold collisions. Jet quenching and suppression of 
leading particles are signs of QGP formation.

Eventually, with plenty of data in hand, all the 
four detector collaborations - STAR, PHENIX, BRAHMS, PHOBOS
 - operating at the BNL have converged on a consensus 
opinion that the fireball is a liquid of strongly 
interacting quarks and gluons rather than a gas of weakly 
interacting quarks and gluons. Moreover, this liquid is 
almost a ``perfect`` liquid with very low viscosity. The 
RHIC findings were reported at the meeting of the American 
Physical Society (APS) held during April 16-19, 2005 in 
Tampa, Florida in a talk delivered by Gary Westfall. Thus, 
it is obvious that the existence of QGP theoretically 
predicted by Quantum Chromodynamics (QCD) has been 
experimentally validated at RHIC.

But the QGP created hitherto in terrestrial 
laboratories is ephemeral, its lifetime is, as mentioned 
earlier, a few times $ 10^{-24} $ second, presumably 
because its temperature is not well above the transition 
temperature for transition from the hadronic phase to the QGP
phase. In addition to this, it is difficult to maintain it 
even at that temperature for long enough time. However, as 
shown in the sequel, in nature we have celestial 
laboratories in the form of gravitationally collapsing 
black holes wherein QGP is created naturally; this QGP is 
at much higher temperature than the transition temperature,
and presumably therefore it is not ephemeral. More so, 
because the temperature of the QGP created in black holes 
continually increases and as such it is always above the 
transition temperature.
\section{Gravitationally collapsing black hole as a particle accelerator }

We consider a gravitationally collapsing black hole
(BH). In the simplest treatment \cite{Weinberg} a BH is considered to 
be a spherically symmetric ball of dust with negligible 
pressure, uniform density $ \rho = \rho(t) $, and at rest at $t=0$. These assumptions lead to the unique solution of the 
Einstein field equations, and in the comoving co-ordinate 
system the metric inside the BH is given by
\begin{eqnarray}
ds^2 = dt^2 -R^2(t)\left[  \frac{dr^2}{1-k\,r^2} + r^2 d\theta^2 + r^2\sin^2\theta\,d\phi^2 \right] 
\end{eqnarray}
in units in which the speed of light in vacuum, $ c=1 $, 
and where $ k=8\pi G \rho(0) /3$ is a constant.
 
On neglecting mutual interactions the energy E of 
any one of the particles comprising the matter in the BH is given by 
$ E^2 = p^2 + m^2 > p^2 $, in units in which again $ c=1 $, and where $p$ is
the magnitude of the 3-momentum of the particle and $m$ its 
rest mass. But $ p = \frac{h}{\lambda}$, where $ \lambda $ is the de Broglie wavelength of the particle and $h$ Planck's constant of action. Since all length in the collapsing BH scale down in proportion to the scale factor $ R(t) $ in equation (1), it is obvious that $ \lambda \propto R(t) $. Therefore it follows that $ p \propto R^{-1}(t) $, and hence $ p = a R^{-1}(t) $, where $a$ is the constant of 
proportionality. From this it follows that $ E > a/R(t)$. 
Consequently, $E$ as well as $p$ increases continually as $R$ decreases. It is also obvious that $E$ and $ p \rightarrow \infty $ as $R \rightarrow 0$. Thus, in effect,
 we have an ultrahigh energy particle accelerator,
so far inconceivable in any terrestrial laboratory, in the 
form of a gravitationally collapsing BH, which can, in the 
absence of any physical process inhibiting the collapse, 
accelerate particles to an arbitrarily high energy and 
momentum without any limit.

What has been concluded above can also be 
demonstrated alternatively, without resorting to the general 
theory of relativity, as follows. As an object collapses 
under self-gravitation, the inter-particle distance $s$ 
between any pair of particles in the object decreases. 
Obviously, the de Broglie wavelength $ \lambda$ of any 
particle in the object is less than or equal to $s$, a simple consequence of Heisenberg's uncertainty principle. 
Therefore, $ s \geq \frac {h}{p}$. Consequently, $ p \geq \frac {h}{s}$ and hence $ E \geq \frac {h}{s} $. Since during the gravitational collapse of an 
object $s$ decreases continually, the energy $E$ as well as $p$, the magnitude of the 3-momentum of each of the particles is 
the object increases continually. Moreover, from $ E \geq \frac {h}{s} $ and $ p \geq \frac {h}{s} $
it follows that $E$ and $ p \rightarrow \infty $ as $s \rightarrow 0 $. Thus, any gravitationally
collapsing object in general, and a BH in particular, acts 
as an ultrahigh energy particle accelerator.

It is also obvious that $ \rho $, the density 
of matter in the BH, continually increases as the BH 
collapses. In fact, $ \rho \propto R^{-3}$, and hence $\rho \rightarrow \infty$ as $ R \rightarrow 0$.
\section{Creation of quark-gluon plasma inside gravitationally collapsing 
black holes}
It has been shown theoretically that when the energy
$E$ of the particles in matter is 
$ \sim 10^{2}$ GeV ($s \sim 10^{-16}$ cm ) corresponding to a temperature $ T \sim 10^{15} $ K, all interactions are of the Yang-Mills type with $SU_{c}(3)\times SU_{I_{W}}(2)\times U_{Y_{W}}(1)$ gauge symmetry, where $c$ 
stands for colour, $ I_{W} $ for weak isospin, and $ Y_{W} $ for 
weak hypercharge; and at this stage quark deconfinement 
occurs as a result of which the matter now consists of its 
fundamental constituents: spin 1/2 leptons, namely, the 
electrons, the muons, the tau leptons, and their neutirnos, 
which interact only through the electroweak interaction; and 
the spin 1/2 quarks, $u$(up), $d$(down), $s$(strange), $c$(charm), $b$(bottom), $t$(top), which interact electroweakly as well as through the colour force generated by gluons \cite{Ramond}. In this 
context it may be noted that, as shown in section 2, the 
energy $E$ of each of the particles comprising the matter in a 
gravitationally collapsing BH continually increases, and so 
does the density $ \rho $ of the matter in the BH. During
the continual collapse of a BH a stage will be reached when
$E$ and $\rho$ will be so large and $s$ so small that the quarks 
confined in the hadrons will be liberated from the \textit{infrared 
slavery} and acquire \textit{asymptotic freedom}, i.e., the quark 
\textit{deconfinement} will occur.This will happen when 
$ E \sim 10^{2}$ GeV ($s \sim 10^{-16}$ cm) corresponding to
 $ T \sim 10^{15} $ K. Consequently, during the continual 
gravitational collapse of a BH, when 
$ E \geq  10^{2}$ GeV ($ s \leq 10^{-16}$ cm) corresponding to $ T \geq 10^{15} $ K, the entire matter in the BH will be in 
the form QGP permeated by leptons.

One may understand what happens eventually to the
matter in a gravitationally collapsing BH in another way as 
follows. As a BH collapses continually, gravitational energy 
is released continually. Since, inter alia, gravitational 
energy so released cannot escape the BH, it will continually 
heat the matter comprising the BH. Consequently, the 
temperature of the matter in the BH will increase 
continually. When the temperature reaches the transition 
temperature for transition from the hadronic phase to the 
QGP phase, which is predicted to be 
$ \sim 170$ MeV ($\sim 10^{12}$ K)  by the Lattice Gauge 
Theory, the entire matter in the BH will be converted into 
QGP permeated by leptons.

It may be noted that in a BH the QGP will not be 
ephemeral like what it has hitherto been in the case of the 
QGP created in terrestrial laboratories, it will not go back 
to the hadronic phase, because the temperature of the matter 
in the BH continually increases and, after crossing the transition temperature for the transition 
from the hadronic phase to the QGP phase, it will be more and
more above the transition temperature. Consequently, once the
transition from the hadronic phase to the QGP phase occurs in
a BH, there is no going back; the entire matter in the BH 
will remain in the form of QGP permeated by leptons.
\section{Conclusion}
From the foregoing it is obvious that a BH acts as an
ultrahigh energy particle accelerator that can accelerate 
particles to energies inconceivable in any terrestrial 
particle accelerator, and that the matter in any 
gravitationally collapsing BH is eventually converted into 
QGP permeated by leptons. However, the snag is that it is not 
possible to probe and study the properties of the QGP in a BH
because nothing can escape outside the event horizon of a BH.
\section{Acknowledgment} 
The author thanks Professor S. K. Pandey, the 
Co-ordinator of the Reference Centre at Pt. Ravishankar 
Shukla University, Raipur of the University Grants 
Commission's Inter-university Centre for Astronomy and 
Astrophysics at Pune. He also thanks Mr. Laxmikant Chaware 
and Miss Leena Madharia for typing the manuscript.



\end{document}